\documentstyle[aps,prl,epsf,twocolumn]{revtex}
\begin{document}
%\preprint
%\widetext
\title{The anharmonic electron-phonon problem}
\author{J.\ K.\ Freericks$^1$, Mark Jarrell$^2$, and G.~D. Mahan$^3$}
\address{$^1$Department of Physics, Georgetown University, Washington, DC 
20057--0995\\
$^2$Department of Physics, University of Cincinnati, Cincinnati, OH 45221\\
$^3$Department of Physics and Astronomy, University of Tennessee, Knoxville,
Tennessee 37996--1200
and Solid State Division, Oak Ridge National Laboratory, P. O. Box 2008,
Oak Ridge, Tennessee 37831--6030\\}
\date{\today}
\maketitle
\begin{abstract}
The anharmonic electron-phonon problem is solved in the infinite-dimensional
limit using quantum Monte Carlo simulation.  Charge-density-wave order
is seen to remain at half filling even though the anharmonicity removes
the particle-hole symmetry (and hence the nesting instability) of the model.
Superconductivity is strongly favored away from half filling (relative to the
charge-density-wave order) but the anharmonicity does not enhance transition
temperatures over the maximal values found in the harmonic limit.
\end{abstract}
\renewcommand{\thefootnote}{\copyright}
\footnotetext{ 1996 by the authors.  Reproduction of this article by any means
is permitted for non-commercial purposes.}
\renewcommand{\thefootnote}{\alpha{footnote}}


\paragraph*{Introduction} Lattice anharmonicity is responsible for many 
different properties of solids.  For example, the expansion of a crystal 
upon heating is due entirely to anharmonic effects---harmonic crystals do not 
change their volume upon heating.  In addition, the phonon contribution to 
thermal resistivity is due, in some measure, to anharmonic effects.
Furthermore, lattice modes in a real crystal are never completely harmonic.  
Anharmonic terms always enter to higher-order when expanding in the amplitude 
of the lattice vibrations.

But the effects of anharmonicity on the electron-phonon problem, and
more specifically on superconductivity, are relatively unknown.  This
is surprising, because the electron-phonon problem is well studied,
especially in the weak coupling limit.  Here the theory of electrons 
interacting with harmonic phonons, initially pioneered by Migdal \cite{migdal} 
and Eliashberg \cite{eliashberg}, has led to an accurate description
of superconductivity in most low-temperature phonon-mediated superconductors.
It has been found, that in the regime where the electron-phonon coupling
is weak so that the Migdal-Eliashberg theory still applies to the anharmonic
case, the superconducting transition temperature is not enhanced by the
presence of lattice anharmonicity \cite{sofo_mahan}.  The situation as
the coupling is increased is less clear.  Recently, a new mechanism for 
superconductivity, driven by kinetic-energy effects, has been hypothesized 
in anharmonic crystals that display stronger electron-phonon coupling 
\cite{hirsch_anharm}.  In this new mechanism, paired holes, in a nearly 
filled band, are more mobile than single holes, driving the superconductivity.

An exact treatment of lattice anharmonicity is, however, difficult from a 
theoretical point of view because an anharmonic ``perturbation''
is never small---the phonon wavefunctions are always dominated by
the anharmonic terms in the potential as the phonon coordinate becomes 
large.  All that is known rigorously about the anharmonic
electron-phonon problem is that the ground state must contain a spin
singlet \cite{freericks_lieb} for even numbers of electrons on a finite lattice.

Much progress can be made in the limit of large spatial dimension
\cite{metzner_vollhardt} where the lattice many-body problem can be mapped
onto a self-consistently embedded impurity problem that is solved via
a quantum Monte Carlo (QMC) simulation.

\paragraph*{Model}
The simplest electron-phonon model that includes anharmonic effects
is the anharmonic Holstein model \cite{holstein} in which the conduction 
electrons interact with local phonon modes:
\begin{eqnarray}
H&=&-  \sum_{i,j,\sigma} t_{ij} c_{i\sigma}^{\dag }c_{j\sigma}
+ \sum_i ( g x_i - \mu )(n_{i\uparrow}+n_{i\downarrow}) \cr
&+&\frac{1}{2M}\sum_i p_{i}^2+\frac{1}{2}M\Omega^2\sum_ix_i^2+\alpha_{anh}\sum_i
x_i^4\quad ,
\label{eq: ham}
\end{eqnarray}
where $c_{i\sigma}^{\dag}$ ($c_{i\sigma}$) creates (destroys) an electron at
site $i$ with spin $\sigma$, $n_{i\sigma}=c_{i\sigma}^{\dag}c_{i\sigma}$ is
the electron number operator, and $x_i$ ($p_i$) is the phonon coordinate
(momentum) at site $i$.  The hopping of electrons is restricted to
nearest-neighbor lattice sites $i$ and $j$ and is governed by the hopping 
matrix element $t_{ij}$ ($t_{ij}$ is a Hermitian matrix).

The local phonon has a mass $M$ and a frequency $\Omega$ associated with 
it.  The anharmonic contribution to the phonon potential energy is chosen 
to be a quartic in the phonon coordinate with a strength $\alpha_{anh}$.  
The deformation potential (electron-phonon interaction strength) is
parameterized by an energy per unit length and is denoted $g$.  A useful
combination of fundamental parameters is the bipolaron binding energy
(in the harmonic limit) $U := -g^2/M\Omega^2$ which determines the energy 
scale for the effective electron-electron interaction mediated by the phonons. 
The chemical potential is $\mu$.

The hopping $t_{ij}$ is isotropic with strength $t=:t^*/2\sqrt{d}$
where $t^*=1$ to define the energy scale and $d$ is the dimensionality.
The mass is then set equal to one $(M=1)$ leaving $U$, $\alpha_{anh}$, 
$\Omega$, and the electron filling as free parameters.  We set $\Omega=0.5t^*$
which is approximately an order of magnitude smaller than the effective
bandwidth $(W\approx 4t^*)$, and we set $U=-4t^*$ 
$(g=1)$ which is well into the strongly coupled regime in the harmonic limit.

\paragraph*{Results}
The original Holstein Hamiltonian corresponds to the case $\alpha_{anh}=0$.
Both the harmonic Holstein model and the harmonic Holstein-Hubbard model
have been solved exactly in the limit of infinite dimensions via quantum
Monte Carlo simulation\cite{freericks_qmc_holst,freericks_qmc_holsthubb}.
These models display charge-density-wave (CDW) order near half filling and 
superconductivity (SC) away from half-filling.  As the phonon frequency is 
increased, the SC is favored relative to the CDW order.  However, in the
strong-coupling limit, CDW order is favored over SC because of the 
band-narrowing effect of the bipolaron.  The quantum Monte Carlo simulations
also found that the effective phonon potential (determined after integrating
out the effects of the electrons) generically acquires a double-well structure
signifying the high-temperature formation of a bipolaron.

An initial analysis of the anharmonic model can be made by employing the 
Born-Oppenheimer approximation in the strong-coupling limit $(t_{ij}=0)$
\cite{hirsch_anharm}:  the phonon frequency is assumed to be smaller than any
of the other energy scales so the phonons can be approximated by static lattice
distortions corresponding to the minimum of the local phonon potential energy.
Since the phonons couple linearly to the electronic charge,
the equilibrium phonon coordinate varies when there are zero, one, or two 
electrons on a site.  The origin $x_0=0$ is chosen to correspond to the case
with no electrons  on a site.  Then $x_1$ and $x_2$
denote the equilibrium coordinates with one or two electrons on a lattice site.
In the harmonic case the relative distances $x_1-x_0$ and $x_2-x_1$ are 
identical, which is a requirement for particle-hole symmetry.  When a
lattice anharmonicity is turned on, the equilibrium phonon coordinates with
one and two electrons on a lattice site all move toward the origin, but the
relative distances are no longer symmetric, rather the distance $x_2-x_1$ 
becomes significantly smaller than $x_1-x_0$, as can be seen in Figure~1(a).

The lattice anharmonicity and subsequent asymmetric shift of the equilibrium 
phonon coordinate have two main effects: (1) the model loses particle-hole 
symmetry which may allow a new type of kinetic-energy driven superconductivity 
to emerge \cite{hirsch_anharm} and (2) the effective electron-electron 
attraction is dramatically reduced as can be seen by a plot of the bipolaron 
binding energy in Figure~1(b).  Thus the lattice anharmonicity generates an 
effective {\it retarded} repulsive interaction between the electrons.  It 
is possible that this effective repulsive interaction between the 
electrons is sufficiently strong to destroy the kinetic-energy driven 
superconductivity.   The lattice anharmonicity also breaks particle-hole 
symmetry, removing the nesting instability of the CDW at half filling and 
weak coupling.  Hence, in the weak-coupling limit, one expects the lattice 
anharmonicity to favor SC relative to CDW order, possibly completely
suppressing CDW order, although it is also likely that anharmonic effects 
will reduce all transition temperatures.  What is surprising is that
a lattice anharmonicity as small as $\alpha_{anh}=0.01$ (cf.~Fig.~1(b)) can 
have such a large effect on the electron-phonon problem in the strongly
coupled regime studied here.

The anharmonic Holstein model is solved exactly in the infinite-dimensional
limit via QMC simulation.  The algorithm is straightforward, and has been
described numerous times before 
\cite{jarrell_hubbard,freericks_qmc_holst,freericks_qmc_holsthubb}, 
so it will not be repeated here.  The QMC
algorithm determines the single and two-particle Green functions at each
of the Matsubara frequencies, which allows one to construct directly the 
momentum-dependent susceptibility to either CDW or SC order of the 
infinite-dimensional lattice.  Divergence of either of these susceptibilities
signals a transition to an ordered state.

We find that at half filling the system orders in a commensurate CDW phase.
The results of the QMC simulation are presented in Figure~2.  The solid dots
represent the QMC transition temperatures, and the dashed line is a guide
to the eye.  The solid line is the result of a strong-coupling perturbation
theory expansion \cite{freericks_strong}
(fourth-order perturbation theory in $t^*$) that predicts that
the system evolves from CDW order (solid line) to SC order (dotted line).
We find no evidence for such an evolution in the QMC solution.  Rather,
the system remains in the CDW-ordered state even though there is
no particle-hole symmetry remaining in the model.  It is possible that as
the strength of the anharmonicity is increased further that the SC order
may overtake the CDW order, but it would be difficult to see that behavior
with the QMC simulations because the transition temperatures become too
small to be reliably determined.  Note that the strong-coupling perturbation
theory is accurate in predicting the peak position and peak height for the
maximal CDW transition temperature.  This is similar to what happened in the
harmonic case as a function of the coupling strength.

The initial rise in $T_c$ occurs because of the reduction in the bipolaron
binding energy $E_{bip}$ as the strength of the anharmonicity increases.
In the strong-coupling regime, the transition temperature is inversely
proportional to $E_{bip}$ ($T_c\propto t^{*2}/E_{bip}$).  As the bipolaron
unbinds, the system evolves from preformed-pair physics to BCS-type
physics where pair formation and condensation both occur at $T_c$.  This
then causes the transition temperature to drop as $\alpha_{anh}$ is increased
further.

This physics can be dramatically illustrated by extracting the effective 
phonon potential from the QMC simulation \cite{freericks_qmc_holst}.
The probability distribution for the local phonon coordinate $x$ is
measured for each of the imaginary time slices.  This probability 
distribution is then equated to an activated form for an effective potential
$P(x)=\exp[-V_{eff}(x)/T]$.  The effective phonon potential includes the
effects of the interaction of the electrons with the phonons.  In the
harmonic limit, this effective potential was seen to evolve from a harmonic
form $(x-x_1)^2$ to a symmetric double-well form as the coupling strength 
increased\cite{freericks_qmc_holst}.  The double well signified bipolaron
formation, since the number of electrons sitting at the lattice site 
was either zero or two for each of the minima (centered around $x_0$ and 
$x_2$).   The results for the effective phonon potential in the anharmonic 
model at half filling and $T=0.125t^*$ are presented in Figure~3.  The 
effective potential is plotted versus a rescaled phonon coordinate for 
different values of $\alpha_{anh}$. The zeroes of the potentials are shifted 
vertically to separate the different plots.  Note that the addition of 
anharmonicity squeezes the phonon coordinate toward the origin, and makes 
the effective phonon potential asymmetric (corresponding to the loss of 
particle-hole symmetry).  The barrier height decreases as $\alpha_{anh}$
increases, as required by the reduction of the bipolaron binding energy.
Note that the maximum CDW $T_c$ occurs when the barrier height is on the order
of $T_c$ ($\alpha_{anh}\approx 0.0075$).

In order to find SC solutions, we must dope the system off half filling.
The phase diagram for three different values of $\alpha_{anh}$ is shown
in Figure~4.  In Figure~4(a), a small value of $\alpha_{anh}=0.005$ is
chosen.  This result is close to the harmonic result.  Nevertheless, 
even at this small value of $\alpha_{anh}$, the phase diagram is not 
symmetric with respect to reflection about half filling, signifying 
the particle-hole asymmetry. We compare to the strong-coupling perturbation 
theory, which predicts CDW order to survive until doped well off of half filling.  This approximation is, by construction, particle-hole symmetric.  
Note that the case of less than half filling is approximated much 
better by the perturbative calculation, than the greater than half-filling 
case.  

Figure~4(b), has $\alpha_{anh}=0.02$ which lies at a point beyond
where the CDW $T_c$ is maximal.  The perturbative approximation is less 
accurate here, but displays the correct order-of-magnitude estimate for
the phase diagram.  We see how the anharmonicity favors the superconducting 
solutions, pushing the CDW-SC phase boundary toward half filling.  SC is
favored because the squeezing of the phonon coordinate toward the origin
enhances the Frank-Condon overlap factors associated with electronic 
hopping and increases the electronic bandwidth.

Figure~4(c), takes the large value $\alpha_{anh}=0.1$.  The strong-coupling 
theory fails in this regime, and so it has not been included in the plot.  
We see once again how the increase in $\alpha_{anh}$ now suppresses the 
transition temperatures, and how it moves the CDW-SC boundary closer to half 
filling.  There is no evidence for the new SC mechanism.

We found no evidence for incommensurate CDW order, either at half filling 
or away from half filling.  The CDW susceptibility still had it's maximum
at the $(\pi,\pi,\pi,\ldots)$ point when the SC susceptibility diverged
at a higher $T_c$.  Anharmonicity does not favor incommensurate order
in this regime of parameter space.

The effect of anharmonicity on the electron-phonon problem has some aspects
that are similar to, and some aspects that are different from the effect of
a direct Coulomb repulsion \cite{freericks_qmc_holsthubb}.  In both cases 
the superconducting order is favored by the additional interaction relative 
to the charge-density-wave order, in the sense that the 
SC-CDW phase boundary moves toward half filling.  However, the anharmonicity
breaks the particle-hole symmetry, while the Coulomb repulsion does not.
Furthermore, the CDW transition temperature is moderately suppressed even 
for large values of the anharmonicity.  Whereas, if the Coulomb repulsion
is increased to large enough values, the CDW $T_c$ will plummet to zero.

\paragraph*{Conclusion}
We have analyzed the effect of anharmonicity on the 
electron-phonon problem in the regime where the harmonic model has formed
bipolarons.  We find that the anharmonicity breaks particle-hole symmetry
and produces phase diagrams that are not symmetric about half filling.
The anharmonicity acts in some respects as a retarded ``Coulomb repulsion'',
since it greatly favors SC solutions relative to CDW ones.  However, in the
infinite-dimensional limit, we
did not find any enhancement to $T_c$'s for the anharmonic model relative
to the maximal values found in the harmonic case.  Furthermore, we
found no evidence for any kind of novel superconducting mechanism driven
by kinetic-energy effects.  We also studied the effective phonon potential
which includes the effects of the interaction of the phonon with the electrons.
We discovered that the anharmonicity tends to squeeze the phonon coordinates 
closer to the origin, and rapidly suppresses the bipolaron formation. We 
also verified that the strong-coupling perturbation theory accurately 
describes the anharmonic Holstein model in the regime where the system still 
has preformed pairs.

We would like to acknowledge useful conversations with W.\ Chung, J.\ Hirsch 
and J.\ Serene.
J.~K.~F. acknowledges the Donors of The Petroleum Research Fund, administered 
by the American Chemical Society, for partial support of this research 
(ACS-PRF\# 29623-GB6) and an Oak Ridge Associated University Junior Faculty 
Enhancement Award for partial support of this research. 
M.~J. acknowledges support of NSF grants DMR-9406678 and DMR-9357199.
G.~D.~M. acknowledges support by the University of Tennessee, 
and by Oak Ridge National Laboratory, managed by 
Lockheed Martin Energy Research Corporation for the
U.S. Department of Energy under contract number
DE-AC05-96OR22464.  This research was also supported in part by grant number
DMR950010P from the Pittsburgh Supercomputing Center, sponsored by the
National Science Foundation.

%\end{document}

\begin{figure}[t]
\caption{(a) Phonon coordinates at the minima of the anharmonic phonon 
potential with zero ($x_0$), one $(x_1)$, and two ($x_2$) electrons on a site 
as a function of $\alpha_{anh}$ in the atomic limit $(t_{ij}=0)$.
(b) Reduction of the bipolaron binding energy as a function of $\alpha_{anh}$
in the atomic limit $(t_{ij}=0)$.}
\end{figure}

\begin{figure}[t]
\caption{Charge-density-wave transition temperature at half filling
as a function of $\alpha_{anh}$.  The QMC results (dots--no SC transitions
were found at half filling with the QMC) are compared to the 
strong-coupling perturbation 
theory for CDW order (solid line) and SC order (dotted line).  The dashed 
line through the QMC points is a guide to the eye.}
\end{figure}

\begin{figure}[t]
\caption{Effective phonon potential for the anharmonic Holstein
model at half filling and $T=0.125t^*$.  The labels on the curves 
are the values of $\alpha_{anh}$.  The
potentials are shifted by adding a constant to them to separate the curves.
}
\end{figure}

\begin{figure}[t]
\caption{Phase diagrams for the anharmonic Holstein model at three
different values of $\alpha_{anh}$: (a) 0.005; (b) 0.02; and (c) 0.1.
The solid dots are CDW phases, the open triangles are SC phases.  The solid
lines are a strong-coupling approximation to the CDW $T_c$, and the dotted 
lines are a strong-coupling approximation to the SC $T_c$.  }
\end{figure}

\begin{figure} 
\epsfxsize=2.5in
\epsffile{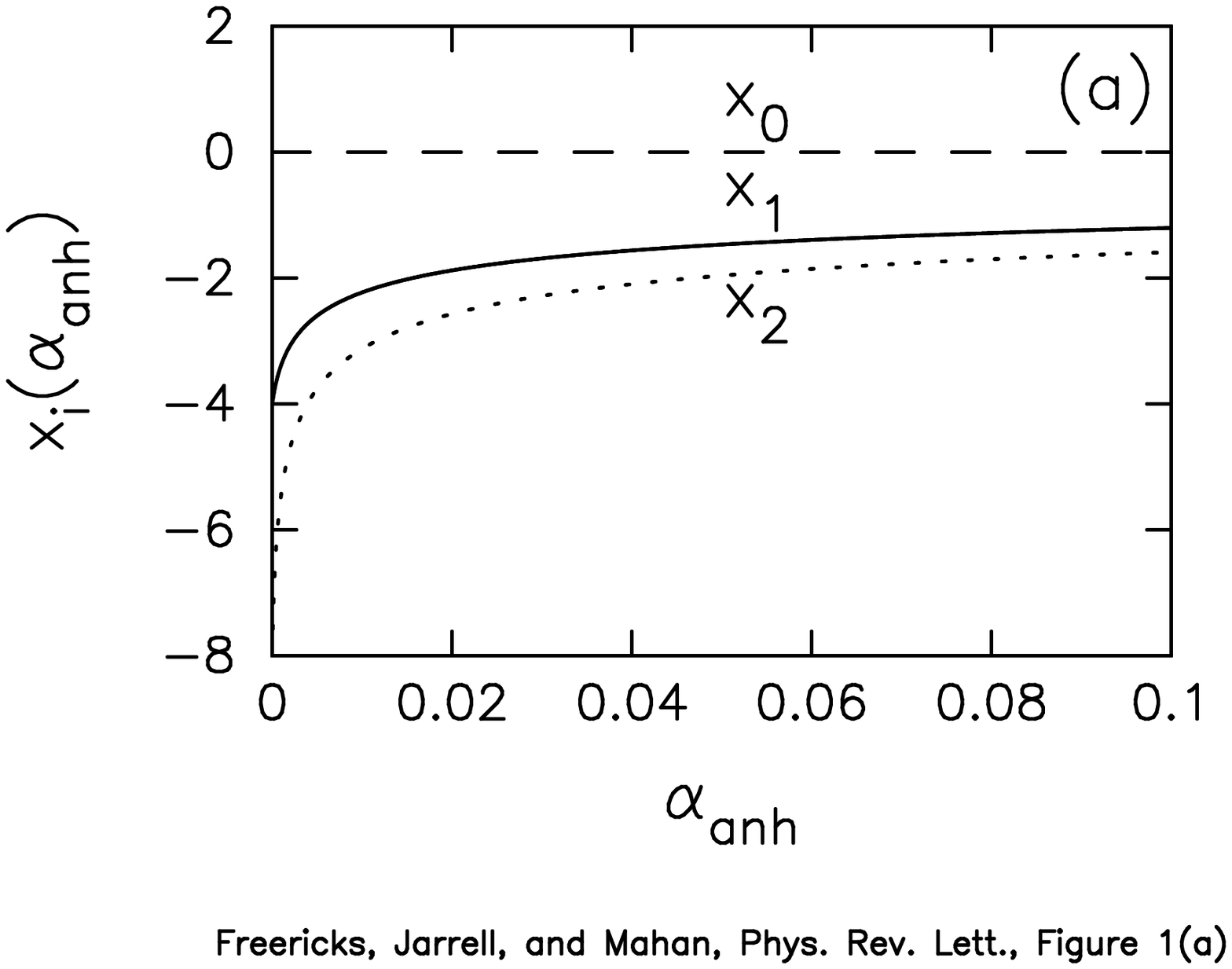}
\end{figure}

\begin{figure}
\epsfxsize=2.5in
\epsffile{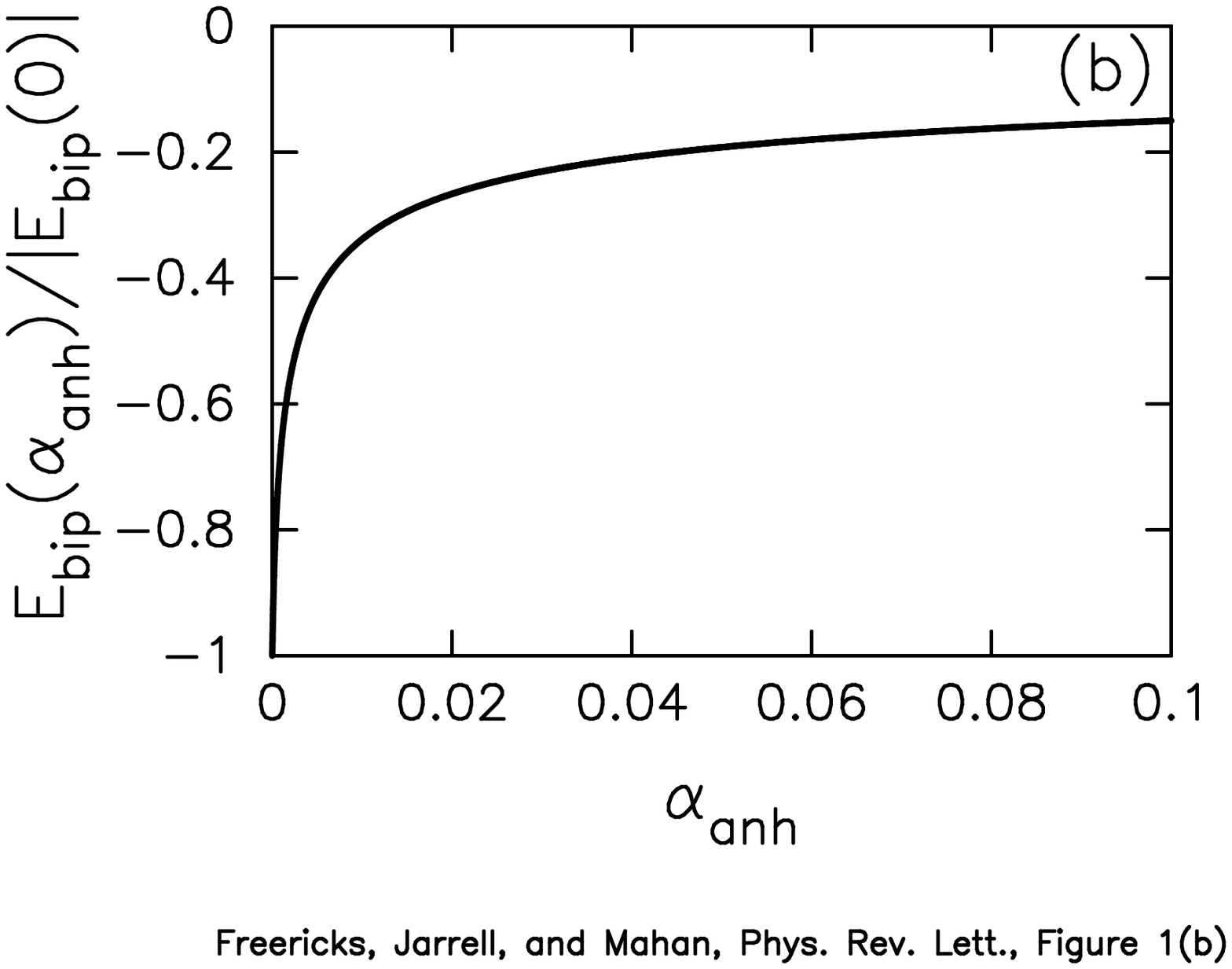}
\end{figure}

\begin{figure}
\epsfxsize=2.5in
\epsffile{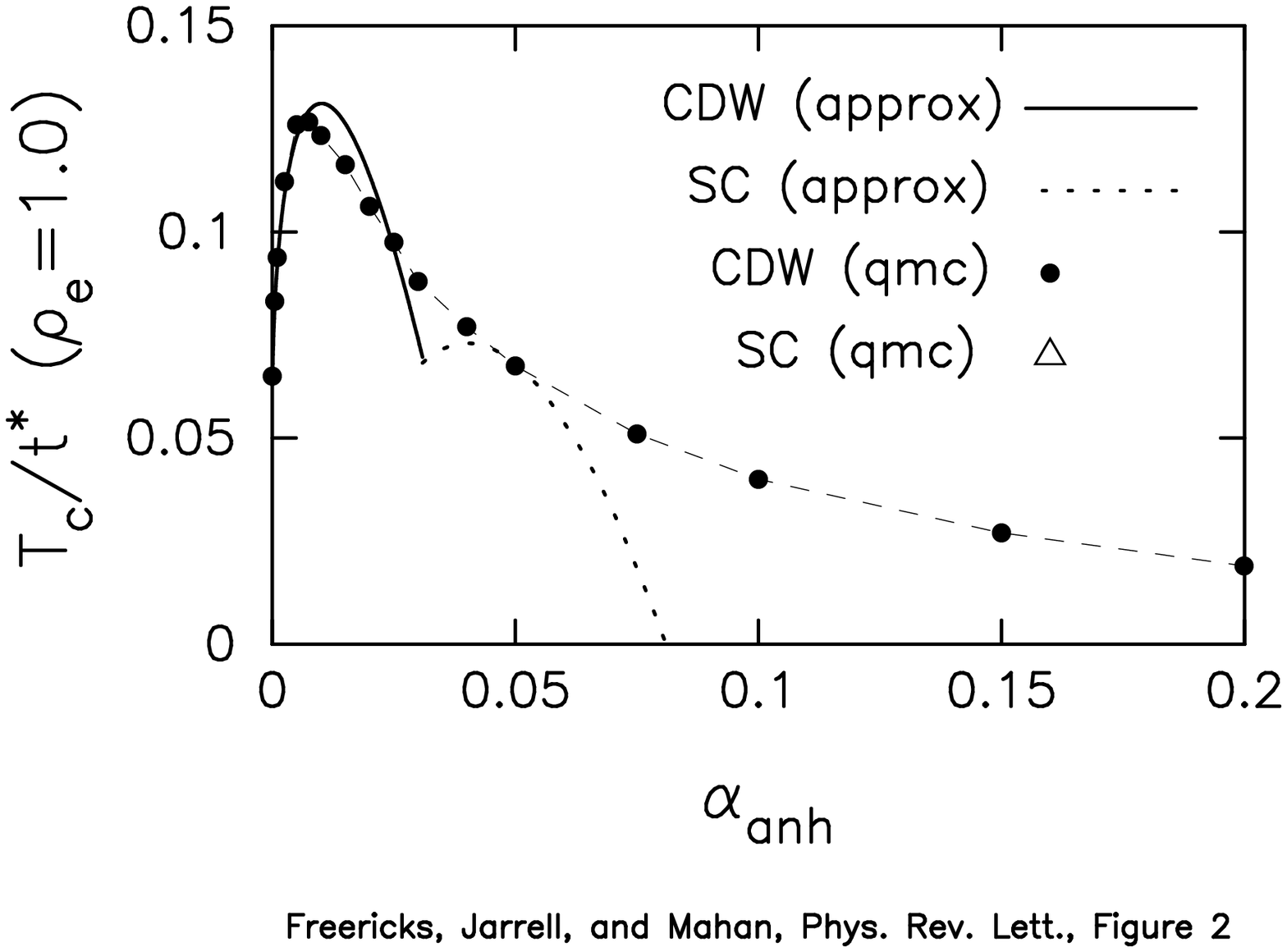}
\end{figure}

\begin{figure}
\epsfxsize=2.5in
\epsffile{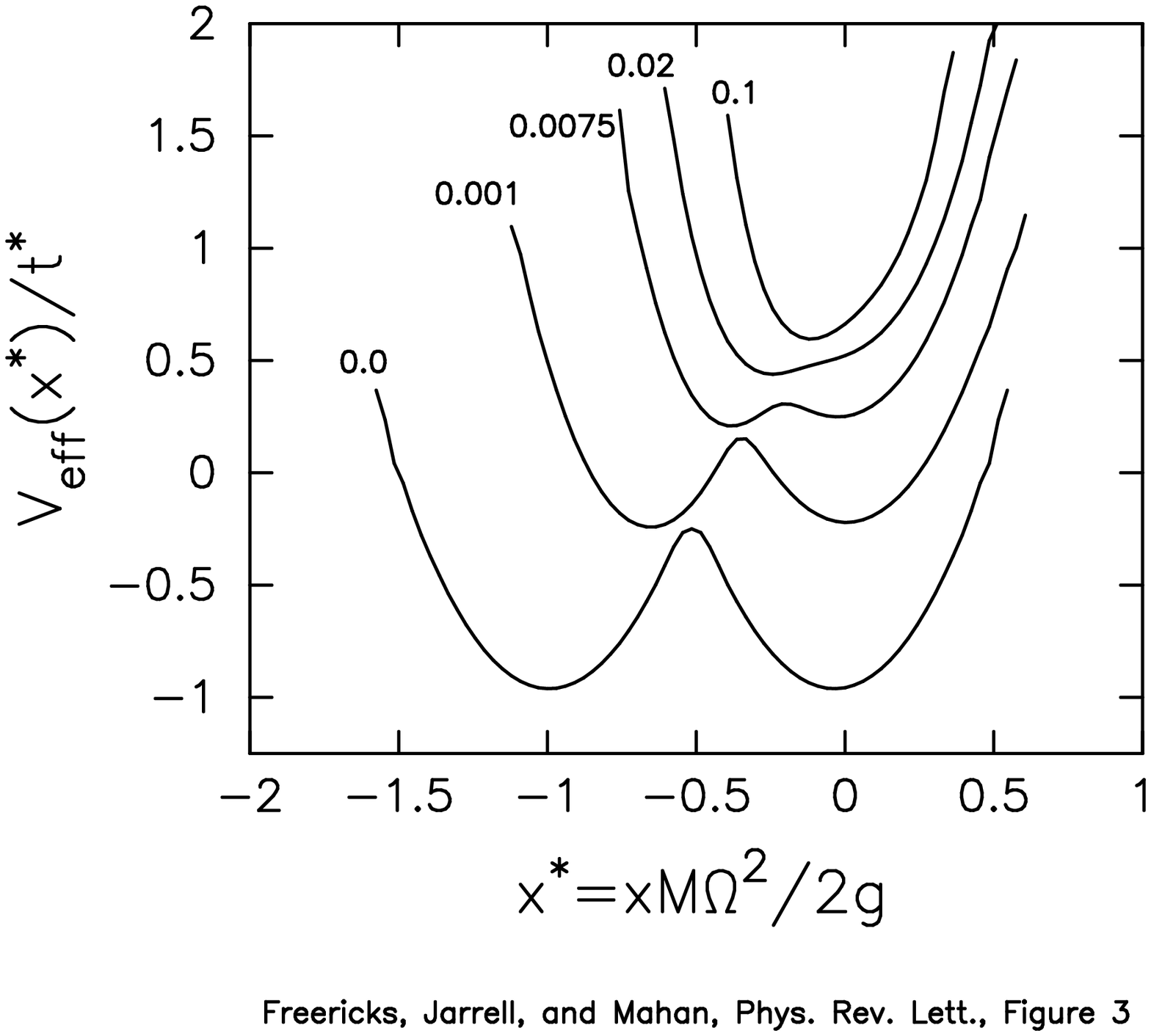}
\end{figure}

\begin{figure}
\epsfxsize=2.5in
\epsffile{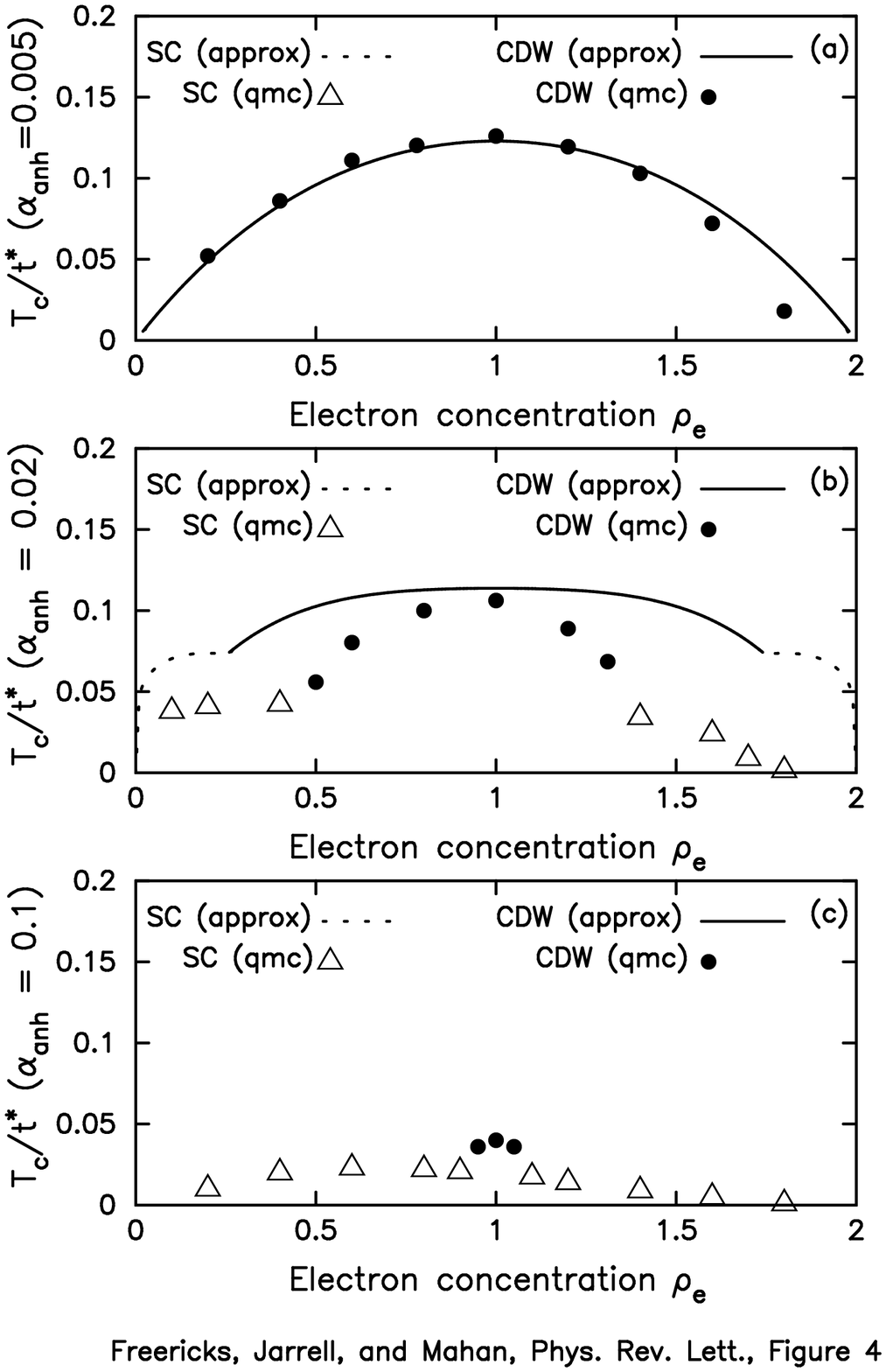}
\end{figure}

\end{document}